\documentclass[twocolumn,showpacs,preprintnumbers,amsmath,amssymb]{revtex4}
\usepackage{graphicx}
\usepackage{epsfig}
\usepackage{dcolumn}
\usepackage{bm}

\newcommand{\NIMA}{Nucl. Instrum. Methods A}

\newcommand{\PLB}{Phys. Lett. B}

\newcommand{\beq}{\begin{equation}}
\newcommand{\eeq}{\end{equation}}
\newcommand{\beqn}{\begin{eqnarray}}
\newcommand{\eeqn}{\end{eqnarray}}
\newcommand{\beqns}{\begin{eqnarray*}}
\newcommand{\eeqns}{\end{eqnarray*}}

\newcommand{\bfg}{\begin{figure}}
\newcommand{\efg}{\end{figure}}
\newcommand{\bitm}{\begin{itemize}}
\newcommand{\eitm}{\end{itemize}}
\newcommand{\bnum}{\begin{enumerate}}
\newcommand{\enum}{\end{enumerate}}
\newcommand{\btbl}{\begin{table}}
\newcommand{\etbl}{\end{table}}
\newcommand{\btbu}{\begin{tabular}}
\newcommand{\etbu}{\end{tabular}}
\begin{document}
\title{\boldmath 
  On Statistical Significance of Signal\footnote{Published in High Ener. 
Phys. Nucl. Phys. 30 (2006) 331. } 
  }
\author{Yongsheng Zhu}
\email{zhuys@ihep.ac.cn}  
\affiliation{ Institute of High Energy Physics, CAS, Beijing 100039, China} 


\begin{abstract}
    A definition for the statistical significance of a signal in an 
experiment is proposed by establishing a correlation between the observed 
$p-$value and the normal distribution integral probability, 
which is suitable for both counting experiment and continuous test statistics.
The explicit expressions to calculate the statistical significance for 
both cases are given.
\end{abstract} 
\maketitle

%
%
\section{Introduction}
The statistical significance of a signal in an experiment of particle 
physics is to quantify the degree of confidence that the observation in 
the experiment either confirm or disprove a null hypothesis $H_0$, 
in favor of an alternative hypothesis $H_1$. Usually the $H_0$ stands for 
a known or background processes, while the alternative hypothesis 
$H_1$ stands for a new or a signal process plus background processes 
with respective production cross section.   This concept is very useful 
for usual measurements that one can have an intuitive estimation, 
to what extent one can believe the observed phenomena are due to 
backgrounds or a signal.  It becomes crucial for the measurements which 
claim a new discovery or a new signal.  As a convention in particle physics 
experiment, the "$5\sigma$" standard, namely the statistical significance 
$S\geq 5$ is required to define the sensitivity for 
discovery; while in the cases  $S\geq 3$ ($S\geq 2$), one may claim that 
the observed signal has strong (weak) evidence.  

However, as pointed out in Ref.~\cite{Sinervo}, the concept of the 
statistical significance has not been employed consistently in the most 
important discoveries made over the last quarter century.  Also, the 
definitions of the statistical significance in different measurements 
differ from each other. Listed below are various definitions for the 
statistical significance in counting experiment (see, for example, 
refs. ~\cite{Bity1}~\cite{Bity2}~\cite{Narsky}):
\begin{equation}
S_1=(n-b)/\sqrt{b},
\end{equation}
\begin{equation}  
S_2=(n-b)/\sqrt{n},
\end{equation}  
\begin{equation}  
S_{12}=\sqrt{n}-\sqrt{b},
\end{equation}   
\begin{equation}    
S_{B1}=S_1-k(\alpha)\sqrt{n/b},
\end{equation}   
\begin{equation}   
S_{B12}=2S_{12}-k(\alpha),
\end{equation}    
\begin{equation}  
\int_{-\infty}^{S_N}N(0,1)dx=\sum_{i=0}^{n-1}e^{-b}\frac{b^i}{i!},
\end{equation}   
where $n$ is the total number of the observed events, which is the Poisson 
variable with the expectation $s+b$, $s$ is the expected number of signal 
events to be searched, while $b$ is the 
known expected number of Poisson distributed background events. All numbers 
are counted in the "signal region" where the searched signal events are 
supposed to appear. In equation (4) and (5), the 
$k(\alpha)$ is a factor related to $\alpha$  that the corresponding 
statistical significance assumes 
$1-\alpha$ acceptance for positive decision about signal observation, and 
$k(0.5)=0,k(0.25)=0.66,k(0.1)=1.28,k(0.05)=1.64$, etc~\cite{Bity2}. 
In equation (6), $N(0,1)$ is a notation for the normal function with 
the expectation and variance equal to 0 and 1, respectively.
On the other hand, the measurements in particle physics often examine 
statistical variables that are continuous in nature. Actually, to identify 
a sample of events enriched in the signal process, it is often 
important to take into account the entire distribution of a given variable 
for a set of events , rather than just to count the events within a given 
signal region of values. In this situation, I. Nasky ~\cite{Narsky} gives 
a definition of the statistical significance via likelihood function
\begin{equation}   
S_{L}=\sqrt{-2\ln L(b)/L(s+b)}
\end{equation}   
under the assumption that $-2\ln L(b)/L(s+b)$ distributes as $\chi^2$ 
function with degree of freedom of 1.

   Upon above situation, it is clear that we desire to have a 
self-consistent definition for statistical significance, which can 
avoid the danger that the same $S$ value in different measurements may 
imply virtually different statistical significance, and can be suitable 
for both counting experiment and continuous test statistics.
In this letter we propose a definition of the statistical significance, 
which could be more close to the desired property stated above.
\section{ Definition of the statistical significance }
The $p-$value is defined to quantify the level 
of agreement between the experimental data and a hypothesis 
Ref.~\cite{Sinervo,PDG}. Assume an 
experiment makes a measurement for test statistic 
$t$ being equal to $t_{obs}$, and $t$ has a probability density function 
$g(t|H_0)$ if a null hypothesis $H_0$ is true. We futher assume that 
large $t$  values correspond to poor agreement 
between the data and the null hypothesis $H_0$, then the 
$p-$value of an experiment would be
\begin{equation}  
           p(t_{obs})=P(t>t_{obs}|H_0)=\int_{t_{obs}}^{\infty}g(t|H_0)dt.
\end{equation}   
A very small $p-$value tends to reject the null hypothesis $H_0$.

   Since the $p-$value of an experiment provides a measure of the 
consistency between the $H_0$ hypothesis and the 
measurement, our definition for statistical significance $S$  
relates with the $p-$value in the form of
\begin{equation}    
         \int_{-S}^{S}N(0,1)dx=1-p(t_{obs})
\end{equation}   
under the assumption that the null hypothesis $H_0$ represents that the 
observed events can be described merely by background processes. 
Because a small $p-$value means a small probability of $H_0$ being 
true, corresponds to a large probability of $H_1$ being true, 
one would get a large signal significance $S$ for a small $p-$value, and 
vice versa. 
The left side of equation (9) represents the integral probability 
of the normal distribution in the region within $\pm S$ standard deviation 
($\pm S\sigma$), therefore, this definition conforms itself to the meaning 
of that the statistical significance should  have.  In 
such a definition, some correlated  $S$  and $p-$values are listed 
in Table~\ref{pS}.
\begin{table}[htbp]
\caption{\it Statistical Significance $S$ and correlated $p-$value.}
\begin{center}
\begin{tabular}{|c|c|}\hline
$S$  &  $p-$value \\\hline
1 & 0.3173  \\
2 & 0.0455  \\
3 & 0.0027  \\
4 & $ 6.3\times 10^{-5} $ \\
5 & $ 5.7\times 10^{-7} $ \\
6 & $ 2.0\times 10^{-9} $ \\\hline
\end{tabular}
\end{center}
\label{pS}
\end{table}

\section{Statistical significance in counting experiment}
A group of particle physics experiment involves the search for new 
phenomena or signal by 
observing a unique class of events that can not be described by 
background processes. One can 
address this problem to that of a "counting experiment", where one 
identifies a class of events 
using well-defined criteria, counts up the number of observed events, 
and estimates the average rate 
of events contributed by various backgrounds in the signal region, 
where the signal events (if exist) 
will be clustered. Assume in an experiment, the number of signal events 
in the signal region is a 
Poisson variable with the expectation $s$, while the number of events from 
backgrounds is a Poisson 
variable with a known expectation $b$ without error, then the observed 
number of events distributes as the Poisson variable with the expectation 
$s+b$. If the experiment 
observed $n_{obs}$ events in the signal region, then the $p-$value is
\begin{eqnarray}
p(n_{obs}) &=&P(n>n_{obs}|H_0)=\sum_{n=n_{obs}}^{\infty}\frac{b^n}{n!}e^{-b} \\
\nonumber
 &=&  1-\sum_{n=0}^{n_{obs}-1}\frac{b^n}{n!}e^{-b} .
\end{eqnarray}
Substituting this relation to equation (9), one immediately has
\begin{equation}   
  \int_{-S}^{S}N(0,1)dx=\sum_{n=0}^{n_{obs}-1}\frac{b^n}{n!}e^{-b}.
\end{equation}   
Then, the signal significance $S$ can be easily determined. Comparing 
this equation with 
equation (6) given by Ref.~\cite{Narsky}, we notice the lower limit of 
the integral is different. 
\section{Statistical significance in continuous test statistics}
The general problem in this situation can be addressed as follows. Suppose 
we identify a class of events using well-defined criteria, which are 
characterized by a set of $N$ observations $X_1,X_2,бн,X_N$ for 
a random variable $X$. In addition, one has a hypothesis to test that 
predicts the probability density function of $X$, say 
$f(X|\vec{\theta})$, where $\vec{\theta}=(\theta_1,\theta_2,...,\theta_k)$ 
is a set of parameters which need to be estimated from the data. Then the 
problem is to define a statistic that gives a measure of the consistency 
between the distribution of data and the distribution given by the 
hypothesis.

To be concrete, we consider the random variable $X$ is, say, an invariant 
mass, and the $N$ observations $X_1,X_2,...,X_N$ give an experimental 
distribution of $X$. Assuming parameters $\vec{\theta}=
(\theta_1,\theta_2,...,\theta_k)\equiv(\vec{\theta_s};\vec{\theta_b)}$, 
where $\vec{\theta_s}$ and $\vec{\theta_b}$ represent the parameters 
related to signal (say, a resonance) and backgrounds 
contribution, respectively.  We assume  the null hypothesis $H_0$ stands 
for that the experimental distribution of $X$ can be described merely by 
the background processes, 
while the alternative hypothesis $H_1$ stands for that the experimental 
distribution of  $X$ should be described by the backgrounds plus signal;
namely, the null hypothesis $H_0$ specifies 
fixed value(s) for a subset of parameters $\vec{\theta_s}$
(the number of fixed parameter(s) is denoted as r), while the alternative 
hypothesis $H_1$ leaves the r parameter(s) free to take any value(s) 
other than those specified in $H_0$. Therefore, 
the parameters $\vec{\theta}$ are restricted to lie in a subspace $\omega$ 
of its total space $\Omega$. On the basis of a data sample of size $N$ 
from  $f(X|\vec{\theta})$ we want to test the 
hypothesis $H_0:\vec{\theta}$ belongs to $\omega$. Given the observations 
$X_1,X_2,бн,X_N$, the likelihood function is 
$L=\prod_{i=1}^{N} f(X_i|\vec{\theta})$. The maximum of this 
function over the total space $\Omega$ is denoted by $L(\hat{\Omega})$; 
while within the subspace $\omega$ the maximum of the likelihood function 
is denoted by  $L(\hat{\omega})$, 
then we define the likelihood-ratio 
$\lambda\equiv L(\hat{\omega})/L(\hat{\Omega})$. It can be 
shown that for $H_0$ true, the statistic
\begin{equation}   
 t \equiv -2\ln \lambda \equiv 2(\ln L_{max}(s+b)-\ln L_{max}(b))  
\end{equation}    
is distributed as $\chi^2(r)$  when the sample size $N$ 
is large~\cite{Eadie}.  In equation (12) 
we use $L_{max}(s+b)$ and $L_{max}(b)$ denoting  $L(\hat{\Omega})$ 
and $L(\hat{\omega})$, respectively.  If $\lambda$ turns out to be in the 
neighborhood of 1, the null hypothesis $H_0$ is such that it renders  
$L(\hat{\omega})$ close to the maximum $L(\hat{\Omega})$, and hence $H_0$ 
will have a large probability of being true. On the other hand, a small 
value of $\lambda$ will indicates that $H_0$ 
is unlikely. Therefore, the critical region of  $\lambda$ is in the 
neighborhood of 0, corresponding to large value of statistic $t$. 
If the measured value of $t$ in an experiment is $t_{obs}$, from 
equation (8) we have $p-$value
\begin{equation}   
 p(t_{obs})=\int_{t_{obs}}^{\infty}\chi^2(t;r)dt.
\end{equation}   
Therefore, in terms of equation (9), one can calculate the signal 
 significance according to following expression:
\begin{equation}    
 \int_{-S}^{S}N(0,1)dx=1-p(t_{obs})=\int_{0}^{t_{obs}}\chi^2(t;r)dt.
\end{equation}    
For the case of $r=1$, we have
\begin{eqnarray*}
 \int_{-S}^{S}N(0,1)dx &=& \int_{0}^{t_{obs}}\chi^2(t;1)dt \\ \nonumber
  &=& 2\int_{0}^{\sqrt{t_{obs}}}N(0,1)dx,
\end{eqnarray*}
and immediately obtain
\begin{eqnarray}
 S &=&\sqrt{t_{obs}} \\ \nonumber
   &=&  [2(\ln L_{max}(s+b)- \ln L_{max}(b))]^{1/2},
\end{eqnarray}
which is identical to the equation (7) given by Ref.~\cite{Narsky}.
\section{Discussion and Summary}
  In section 2, the $p-$value defined by equation (8) is based on the 
assumption that large $t$ values correspond to poor agreement between the 
null hypothesis $H_0$ and the observed data, namely, the critical region 
of 
statistic $t$ for $H_0$ lies on the upper side of its distribution. If the 
situation is such that the critical region of statistic $t$ lies on the 
lower side of its distribution, then equation (8) should be replaced by
\begin{equation}  
p(t_{obs})=P(t<t_{obs}|H_0)=\int_{-\infty}^{t_{obs}}g(t|H_0)dt,
\end{equation}   
and the definition of statistical significance $S$ expressed by equation 
(9) is still applicable.  For the case that the critical region of 
statistic $t$ for $H_0$ lies on both lower and upper tails of its 
distribution, and one determined from an experiment the observed $t$ 
values in both sides: $t_{obs}^U$ and $t_{obs}^L$, then equation (8) 
should be replaced by 
\begin{eqnarray}       
p(t_{obs}) &=& P(t<t_{obs}^L|H_0)+P(t>t_{obs}^U|H_0)   \\ \nonumber
&=& 
\int_{-\infty}^{t_{obs}^L}g(t|H_0)dt+\int_{t_{obs}^U}^{\infty}g(t|H_0)dt.
\end{eqnarray}

  In summary, we proposed a definition for the statistical significance 
by establishing a correlation between the normal distribution 
integral probability and the $p-$value observed in an experiment, which
is suitable for both counting experiment and continuous test statistics.
The explicit expressions to calculate  the statistical significance for 
counting experiment and continuous test statistics in terms of 
the Poisson probability and likelihood-ratio are given.

%

%
\end{document}